\documentclass{article}

\usepackage{geometry}
\usepackage{amsmath,amssymb,amsthm,bm,verbatim}
\usepackage{latexsym,color}
\usepackage{graphicx,amsfonts,natbib}
\usepackage{mycommands}
\usepackage[english]{babel}
\usepackage{authblk}  

\usepackage{amssymb,dsfont}
\usepackage{graphicx,psfrag,epsf}
\usepackage{enumerate}
\usepackage{subcaption}
\usepackage{bbm,rotating}

\newcommand{\bfX}{{\bf X}}
\newcommand{\bfD}{{\bf D}}
\newcommand{\bfU}{{\bf U}}
\newcommand{\bfZ}{{\bf Z}}

\newcommand{\bfA}{{\bf A}}

\newcommand{\bfV}{{\bf V}}
\newcommand{\bfB}{{\bf B}}
\newcommand{\bfH}{{\bf H}}
\newcommand{\bfG}{{\bf G}}

\newcommand{\bfv}{{\bf v}}
\newcommand{\bfu}{{\bf u}}
\newcommand{\bfz}{{\bf z}}
\newcommand{\bfr}{{\bf r}}
\newcommand{\bfc}{{\bf c}}
\newcommand{\bfp}{{\bf p}}
\newcommand{\bfI}{{\bf \mathbb{I}}}
\newcommand{\bfone}{{\mathds{1}}}
\newcommand{\centre}{{\left(\bfI_n-\frac{1}{n}\bfone\bfone^{T}\right)}}

\newtheorem{theorem}{Theorem}

\newtheorem{lemma}[theorem]{Lemma}

\begin{document}

\title{Multiple Correspondence Analysis \& the Multilogit Bilinear Model}

\author[1]{William Fithian}
\author[2]{Julie Josse}
\affil[1]{\small Berkeley Statistics Department, USA}
\affil[2]{\small  Agrocampus Ouest - INRIA, Saclay, France}

\date{}
\maketitle

\begin{abstract}
Multiple Correspondence Analysis
  (MCA)  is a dimension reduction method which plays a large role in the analysis of
  tables with categorical nominal variables such as survey data.  Though it is usually motivated and derived
  using geometric considerations, in fact we prove that it amounts to a single proximal Newtown step of a natural bilinear
  exponential family model for categorical data the {\em multinomial logit bilinear model}. 
We compare and contrast the behavior of MCA with that of the model on simulations and discuss new insights on the properties of both exploratory multivariate methods and their cognate models. One main conclusion is that we could recommend to approximate the \textit{multilogit model} parameters
using MCA. Indeed, estimating the parameters of the model is not a trivial task whereas MCA has the great advantage of being easily solved by singular value decomposition and scalable to large data.  


keywords: nominal data,  dimension reduction, low-rank approximation, latent-space models, contingency table, correspondence analysis.
\end{abstract}

\section{Introduction}

Principal component methods such as principal component analysis
(PCA), correspondence analysis (CA) \citep{Green07} or multiple correspondence
analysis (MCA) \citep{Green06}  are often used as multivariate descriptive methods to
explore and visualize data. They are similar in
their main aims but differ with respect to the nature of the data they deal with: principal component analysis for quantitative data, correspondence analysis for contingency tables (crossing two categorical variables) and multiple correspondence analysis for categorical data.  These data dimensionality reduction methods allow to study the similarities between rows, similarities between columns, and associations between rows and columns and provide a subspace that best represents the data in the sense of maximizing the variability of the projected points. A great importance is attached to the graphical outputs to shed lights into the results and often the representation of rows is as interesting as the columns one \citep{Paghuss10}.

An intrinsic characteristic of the approaches is that they are 
motivated by geometrical considerations without any reference
to probabilistic models, in line with \cite{Benz73}'s idea to ``let the
data speak for themselves''.  From a technical point of view, the core of all these methods is the singular value decomposition (SVD) of certain matrices with specific row and column weights and metrics (used to compute the distances).  In the words of \cite{Benz73}, ``Doing a data analysis, in
good mathematics, is simply searching eigenvectors, all the science of
it (the art) is just to find the right matrix to diagonalize.''

Even so, specific choices of weights and metrics can be viewed as
inducing specific models for the data under analysis.  Understanding
the connections between exploratory multivariate methods and
their cognate models can yield insights into the methods' properties
and offer for instance solutions when explicit models struggle with high dimensional data. Indeed,  principal components methods have the great advantage to be  easily scalable to large data sets.
In addition, it may give new opportunities to common problems for principal components methods such as inference, tests to select the number of components, and missing values.

In this paper, we begin in Section \ref{sec:pca_ca_models} with a brief review of PCA, CA and their cognate models in one place, with a focus on their similarities. We also include a new presentation of CA as a generalized linear model with a data driven link function. 
Then, we describe in Section \ref{sec:mca_models} the {\em multinomial logit bilinear  model}  to study the structure of dependence between categorical variables
and derive theoretical results relating MCA to this model.   We show that MCA amounts to a single proximal Newton step on the multilogit model. 
Finally, in Section \ref{sec:simulations} we conduct a simulation study to compare and contrast the behavior of the \textit{multilogit model} with that of MCA and discuss the potential of such new connections. 


\section{Underlying Models in PCA and in CA\label{sec:pca_ca_models} }

\subsection{The Linear-Bilinear Model and PCA\label{sec:pca}}

A classical model related to PCA is the fixed-effects model of \cite{Caussinus86},
also known as the fixed factor score model \citep{Leeuw85} and discussed in \cite{gls}.  In that
model, the data matrix $\bfX\in \R^{n\times m}$ is generated from column effects and
a rank-$K$ interaction matrix, corrupted by isotropic Gaussian
noise:
\begin{equation}\label{mod_acp}
  x_{ij} \sim \mathcal{N}(\mu_{ij},\sigma^2), \text{  with  } \mu_{ij} =
  \beta_j + \Gamma_{ij},
\end{equation}
with the constraint that $\text{rank}(\boldsymbol{\Gamma})\leq K$.  Equivalently,
we can write
\begin{equation*}
  \mu_{ij} = \beta_j + \sum_{k=1}^K d_k u_{ik}v_{jk},
\end{equation*}
with identifiability constraint $\bfU^T\bfU = \bfV^T\bfV = \bfI_K$.
Maximum likelihood estimation of $\boldsymbol{\Gamma}$ amounts to least-squares
approximation of the column-centered data matrix $\bfZ =
\centre\bfX$, the matrix of residuals after orthogonalizing
with respect to the
column effect $\boldsymbol{\beta}$.  That is, $\boldsymbol{\widehat  \Gamma}$ is simply the
rank-$K$ partial singular value decomposition (SVD) $\bfU_K \bfD_K \bfV_K^T$, leading to the classical PCA
normalized scores $\bfu_{i}$ and loadings $\bfv_{j}$. The solution can also be obtained using an alternating least squares algorithm (with a multiple regression step to estimate $\bfU$ and a multiple regression step to estimate $\bfV$). 

Model \eqref{mod_acp} is also called a \textit{linear-bilinear
  model} \citep{Mandel69, Denis94, Denis96,
  Groenen06}, an  additive main effects and
multiplicative interaction model (AMMI) or a biadditive model \citep{Gab78, Gauch88,Gow95} and is extremely popular in analysis of
variance with two factors.  In that case one often includes an additive
row effect as well:
\begin{align}
\mu_{ij}&=\alpha_i+\beta_j+\sum_{k=1}^K d_k u_{ik}v_{jk} \label{ammi}
\end{align}
Model \eqref{ammi} is useful to estimate the interaction between the factors when no replication is available. 
\subsection{The Log-Bilinear Model and CA} \label{sec:CA}

Log-linear models \citep{Agresti13, Christensen90} are often used to model counts in contingency tables. The saturated log-linear model for the table $\bfX_{n \times m}$ is:
\begin{equation}
  \log \mu_{ij} = \alpha_i+ \beta_j + \Gamma_{ij} \label{loglin}
\end{equation}
Typically, the $\mu_{ij}$ represent either means of independent Poisson
$x_{ij}$, or the probability of cell $\{ij\}$ in a multinomial
model (\textit{i.e.} obtained by conditioning the Poisson model on the
overall margin $N = \sum_{ij} x_{ij}$).
Although this model is overspecified as written, we can simplify~\eqref{loglin} as in \eqref{ammi} by constraining the rank of the interaction
matrix $\boldsymbol{\Gamma}$:
\begin{equation}
\log \mu_{ij} = \alpha_i+ \beta_j + \sum_{k=1}^K d_k u_{ik} v_{jk} \label{RC_model}
\end{equation}
Model \eqref{RC_model}  is defined by \citet{Good85} as the RC($K$) model (for row-column) and is also called the
\textit{log-bilinear model} \citep{falg98, Gow11c} or GAMMI models (for generalized AMMI). 

Note that the parameters of~\eqref{RC_model} may be interpreted as describing
latent variables in a low-dimensional Euclidean space.  Suppose that
row $i$ of the table corresponds to the point $\tilde \bfu_i = \bfD_K^{1/2}
\bfu_i$, and column $j$ corresponds to $\tilde \bfv_j = \bfD_K^{1/2}\bfv_j$.
Then, we can rewrite~\eqref{RC_model} as
\begin{equation}\label{eq:latent}
  \log \mu_{ij} = \alpha_i + \beta_j + \tilde \bfu_i^T \tilde \bfv_j
  = \tilde \alpha_i + \tilde \beta_j - \frac{1}{2} \|\tilde \bfu_i -
  \tilde \bfv_j\|^2
\end{equation}
That is, $\mu_{ij}$ is large when $\tilde \bfu_i$ and $\tilde \bfv_j$ are
close to each other. Equation~\eqref{eq:latent} is also called a two-mode distance-association model by \citet{Rooij05} and \citet{rooij08}. 


However, solving these low-rank log-linear models is non-trivial: there is no closed-form analog to the partial SVD
outside the context of least-squares estimation  (as  in~\eqref{mod_acp}) and standard methods based on iterative weighted least squares (IWLS), where steps of generalized linear regressions (GLM) are alternated are known to encounter difficulties \citep{fred}. This happens especially when the rank $k$ is greater than 1, the tables are sparse and the total number of counts small.  Maximizing a penalized version of the poisson likelihood \citep{Salmon14} or using a Bayesian approach \citep{Li2013SEPCA} may be useful to tackle the overfitting issues. 

Contrary to PCA (Section \ref{sec:pca}), there is not an exact correspondence between CA and the \textit{log-bilinear model} \eqref{RC_model}. However, they are  closely related.  
CA \citep{Greenacre84,Green07} is a very powerful method that has been successively used to visualize many contingency tables such as texts corpus tables \citep{Leb98} where texts are characterized by their profile of words. Note also that CA underlies variants of many modern machine learning applications such as spectral clustering on graphs \citep[e.g.,][]{ng2002spectral,shi2000normalized} or topic modeling. 
To perform CA on
a two-way table, we first compute the ``correspondence matrix'' by dividing $\bfX$ by its grand total N and then we compute its row margins $\bfr$ and column
margins $\bfc$ to construct the matrix of {\em pseudo-residuals} $\bfZ$ with
\begin{equation*}
  z_{ij} = \frac{x_{ij}/N - r_i c_j}{\sqrt{r_i c_j}}
\end{equation*}
We can alternatively write $\bfZ$ in matrix form as $\bfZ=
\bfD_\bfr^{-1/2}(\bfX/N-\bfr\bfc^T)\bfD_\bfc^{-1/2}$,
with $\bfD_\bfr=\diag(r_1,\ldots,r_n)$ and $\bfD_\bfc=\diag(c_1,\ldots,c_m)$.
Meanwhile, if $\bfX$ is the adjacency matrix of a graph, then $\bfZ$ is a version of the symmetric normalized graph Laplacian where we have projected out the first trivial eigencomponent. 
Note also that $\|\bfZ\|_F^2$ is exactly the Pearson $\chi^2$ statistic for
the row-column independence model $\hat x_{ij}/N = r_i c_j$.  Hence,
each $z_{ij}$ represents the normalized, signed deviation of $x_{ij}$
from that model.  Once we have formed $\bfZ$, we compute its rank-$K$ partial SVD $\bfZ= \widetilde  \bfU_K \bfD_K  \widetilde  \bfV_K^T$.
The CA  standard row coordinates are then defined as $\bfU_K =
\bfD_\bfr^{-1/2}\widetilde  \bfU_K$ and the standard column coordinates as 
$ \bfV_K = \bfD_\bfc^{-1/2}\widetilde \bfV_K$ and used in biplot representation. 

If the low-rank approximation is good,
then we have
\begin{equation*}\label{eq:caApprox1}
\bfU_K \bfD_K  \bfV_K^T \approx \bfD_\bfr^{-1/2}\bfZ \bfD_\bfc^{-1/2} = \bfD_\bfr^{-1}(\bfX/N-\bfr\bfc^T)\bfD_\bfc^{-1},
\end{equation*}
in a weighted least-squares sense.  By ``solving for $\bfX$'' in
\eqref{eq:caApprox1}, we may obtain the {\em reconstruction formula}
below:
\begin{equation*}\label{eq:caRecon1}
\widehat \bfX/N =  \bfr\bfc^T + \bfD_\bfr(\bfU_K \bfD_K \bfV_K^T)\bfD_\bfc
\end{equation*}
or, rewritten elementwise,
\begin{equation}
\frac{\hat x_{ij}}N{}= r_i c_j\left(1+ \sum_{k=1}^K d_k u_{ik} v_{jk}\right)
\label{rec_ca}
\end{equation}

We suggest here an alternative presentation of CA using a  classical generalized linear framework \citep{ned}. With standard notations,  let us consider the expectation as $\mu_{ij} = r_i c_j\left(1+ \sum_{k=1}^K d_k u_{ik} v_{jk}\right) = M_0(1 + \eta_{ij})$. It leads to a link function that is data driven $g(\mu_{ij}) = \eta_{ij}$. 
We can maximize the Gaussian likelihood using iterative weighted least-squares by defining 
$$z_{ij}^{\ell} = (x_{ij} - \mu_{ij}^{\ell}) \times g'(\mu_{ij})+ \eta_{ij}^{\ell} =  {(x_{ij} - \mu_{ij}^{\ell}) }\frac{1}{M_0}+ \eta_{ij}^{\ell}$$ and 
$$w_{ij}^{\ell} = \frac{1}{V(\mu_{ij}^{\ell})g'(\mu_{ij}^{\ell})^2}= M_{0}$$
To estimate the parameters, we alternate two steps of weighted (with weights $M_0$) linear regressions  of $\bfz$ on $\bfU$ and on $\bfV$. We straightforwardly incorporated it in classical softwares such as  the R package \textit{gnm} \citep{gnm} defining a Gaussian data dependent link function (R code is available on the github repository \citet{git}). It leads to run correspondence analysis for two dimensions with the following line of pseudo-code, which easily encourages the introduction of additional variables in CA which could be extremely useful: \\\textit{CA2 $\gets$ gnm(vect(X) \~\ X1+X2+instances(Mult(X1, X2), 2), family = gaussian(CA($M_0$)), weights=1/$M_0$)}\\

Concerning the connection between CA and the \textit{log-bilinear model}, \cite{Escouf82} highlighted that when $\sum_{k=1}^K d_k
u_{ik} v_{jk}$ is small compared to one, \eqref{rec_ca} can be
approximated by:
\begin{eqnarray*}
\log(\hat x_{ij}) \approx \log(N) +\log(r_i) + \log(c_j)+ \sum_{k=1}^K d_k u_{ik} v_{jk} \label{approx_ca_loglin}
\end{eqnarray*}
Consequently, the CA parameters can be seen as providing an approximation
of the \textit{log-bilinear} parameters \eqref{RC_model} when the interaction
is small. However, \cite{Heijden94} showed empirically that even when
the estimated interaction is large,
parameters obtained by CA and by the \textit{log-bilinear model} are often
very close.
\citet{Heij85} and \citet{Heijd89} studied in depth the relationship between
\textit{log-bilinear model} and CA, highlighting the benefits of using both
methods as complementary data analysis tools. We could also note that depending on the point of view, the \textit{log-bilinear model} can also be seen as an approximation of CA. 


\section{Methods for Analyzing Multiple Categorical Variables} \label{sec:mca_models}

We now proceed to describe two
different frameworks for analyzing
categorical data ---  the {\em multinomial logit bilinear model} and MCA.  As we will see in Section~\ref{sec:connections},
the methods are more closely related than meets the eye, since MCA can
be viewed as a one-step estimate for low-rank versions of the model-based method and as far as we know no direct relationship between these models and  MCA
has been yet established.

\subsection{The Multilogit-Bilinear Model}\label{sec:multiLogitBilinear}

When each categorical variable is binary,
\citet{Collins01ageneralization, tapio2002VEM,  Villardon06, Li2013SEPCA} studied a generalization of the
model~\eqref{RC_model}:
\begin{equation}\label{eq:logisticBilinear}
  \P(x_{ij}=1) = \frac{e^{\theta_{ij}}}{1+e^{\theta_{ij}}},
  \text{  with  } \theta_{ij} = \beta_j + \sum_{k=1}^K d_k u_{ik} v_{jk}
\end{equation}
 This model is also a straightforward
extension of the model~\eqref{mod_acp}
with a different link function (the logit) and can be called a
\textit{logit-bilinear model}.  It is a special case of a {\em
  generalized bilinear model} as defined by \citet{Choulakian} and \citet{Gab98}.
\citet{deLeeuw:2006:PCA} suggested a majorization algorithm to estimate the model's parameters. 
Model ~\eqref{eq:logisticBilinear} is also known as a {\em latent
  traits} model \citep{Lazar68}, since the relationship between the $m$ variables arise
through their mutual dependence on $\bfu_i$, individual $i$'s latent
type.
Popular latent traits model include item-response theory (IRT) models \citep{lin97} but most often in IRT applications $K=1$  and the latent parameter is assumed to be Gaussian, whereas we will consider it as a
fixed parameter. 
\citet{hoff_2009_cmot} and  \citet{raftery_niu_hoff_yeung_2010_tr} also considered related random effect models to analyze network data.

For the case with many categories for each variable, and denoting $x_{ij}=c\in
\{1,\ldots,C_j\}$ the $j$th categorical response for individual
$i$, a natural
extension of \eqref{eq:logisticBilinear} is
\begin{equation}\label{eq:multiLogitBilinear}
  \P(x_{ij}=c) = \frac{e^{\theta_{ij}(c)}}{\sum_{c'=1}^{C_j} e^{\theta_{ij}(c')}},
  \text{  with  } \theta_{ij}(c) =  \beta_j(c) + \Gamma_i^j(c)  = \beta_j(c) + \sum_{k=1}^K d_k u_{ik} v_{jk}(c),
\end{equation}
which can be called the
{\em multilogit-bilinear model}.

The interaction $\Gamma = \left[\Gamma^1 \cdots \Gamma^m\right]$ is constrained to 
have rank $K$.  We add the additional
identifiability constraint that $\Gamma^j p^j = 0$, or
$\sum_c\Gamma_i^j(c) p^j(c) = 0$ for each $i,j$.

\begin{figure}
  \centering
  \includegraphics[width=.4\textwidth]{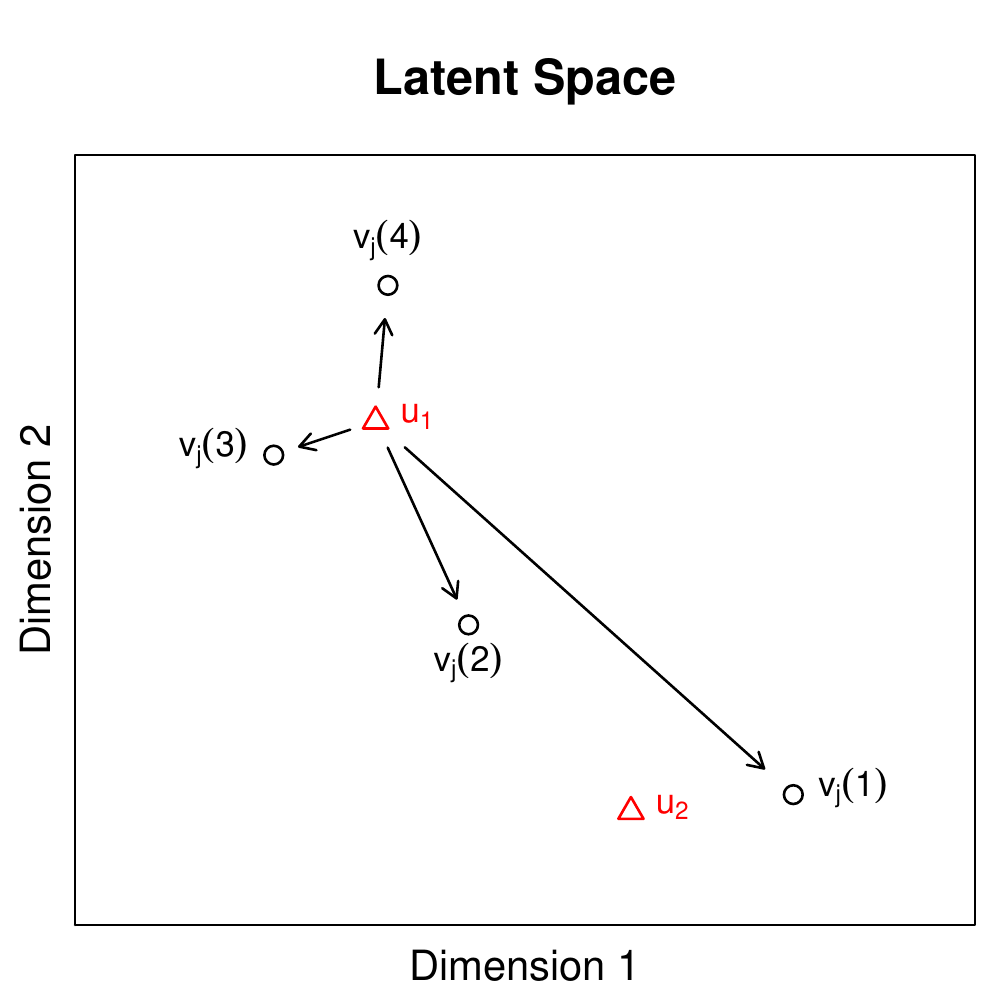}
  \caption{Depiction of the \textit{multilogit-bilinear model} in latent
    space.  For variable $j$,
    individual 1 is more likely to choose response 3 or 4
    than individual 2 is.}
    \label{fig:mlblLatent}
\end{figure}

Though the model~\eqref{eq:logisticBilinear} may seem rather opaque
with its four different indices $i,j,k,$ and $c$, we can show that there is a simple
interpretation of it along the same lines as~\eqref{eq:latent}, which
we depict in Figure~\ref{fig:mlblLatent} for $K=2$.
For question $j$, we associate category $c$ with coordinates
$\tilde \bfv_j(c) = (\sqrt{d_1}v_{j1}(c),\sqrt{d_2}v_{j2}(c))$, yielding one point
for each of the $C_j$ categories.  The latent variables $\tilde \bfu_i = \bfD_{2}^{1/2}\bfu_i$
are plotted for 2 individuals as well.  Then,
\begin{equation*}
  \P(x_{ij} = c) \propto \exp\left\{\tilde\beta_j(c) - \frac{1}{2}\|\tilde \bfv_j(c) - \tilde \bfu_i\|^2\right\}
\end{equation*}
That is, the distribution of $x_{ij}$ depends on the latent-space
distance between the individual and the various categories, as well as
an additional factor depending only on the categories and not the individual.

The estimation task is also non-trivial in part also because of the non-convexity  of the rank constraint. In addition, overfitting issues due to the so-called separability problems inherent of such models cause some of the parameters wandering off to infinity. Consequently, 
\cite{groen} suggested a majorization approach to estimate the parameters but minimized a penalized deviance using the trace norm. 
Note that random-effects version of this model (assuming a Gaussian distribution on the latent variables) has been studied in \cite{mousta0}  who also highlighted the necessity to resort to regularization.  

\subsection{Multiple Correspondence Analysis} \label{sec:mca}

MCA has been successfully applied to visualize the relationship between categorical variables on many fields such as social sciences, marketing, health, psychology, educational research, political science, genetics, etc. \citep{Green06}. MCA is also known as homogeneity analysis or dual scaling  \citep{Mich98,Nishisato80,leeuw14, leroux10, gigi}. 

To characterize MCA, we begin by defining from the data matrix $\bfX$ with $n$ individuals and $m$ variables,  the indicator matrix $\bfA=\left[ \bfA^1 | \cdots |
  \bfA^m\right]$ so that $\bfA^j \in \{0,1\}^{n\times C_j}$, with row $i$
corresponding to a dummy coding of $x_{ij}$.  Alternatively,   with $C = \sum_j C_j$ the total number of levels across all
variables,  define
the so-called Burt matrix $\bfB = \bfA^T \bfA \in \Z^{C
  \times C}$, which contains all two-way tables between pairs
of variables.  Note that $\bfB$ also has a block form,
with $B^{j,j'} = \bfA_j^T \bfA_{j'}$.  For example, with $m=2$ variables
with $C_1=2$ and $C_2=3$ levels respectively, then
\begin{equation*}\label{eq:catTables}
  \bfX = \begin{pmatrix}
    1 & 1\\
    2 & 3\\
    1 & 2\\
    2 & 3\\
    2 & 2\\
    2 & 2
  \end{pmatrix}
  \;\;\; \Longleftrightarrow \;\;\;
  \bfA= \begin{pmatrix}
    1 & 0 & & 1 & 0 & 0\\
    0 & 1 & & 0 & 0 & 1\\
    1 & 0 & & 0 & 1 & 0\\
    0 & 1 & & 0 & 0 & 1\\
    0 & 1 & & 0 & 1 & 0\\
    0 & 1 & & 0 & 1 & 0
  \end{pmatrix}
  \;\;\; \Longrightarrow \;\;\;
  \bfB = \begin{pmatrix}
    2 & 0 & & 1 & 1 & 0\\
    0 & 4 & & 0 & 2 & 2\\
    & & & & & \\
    1 & 0 & & 1 & 0 & 0\\
    1 & 2 & & 0 & 3 & 0\\
    0 & 2 & & 0 & 0 & 2\\
  \end{pmatrix}
\end{equation*}
Note that $\bfX$ and $\bfA$ are equivalent codings of the data, whereas some
information is lost in computing $\bfB$.  Write
$p_j(c)=\frac{1}{n}A_{\cdot c}^j$ for the $c$th normalized
column margin of $\bfA^j$, with $\bfp=(\bfp_1,\ldots,\bfp_m)^T$.
All row margins of $\bfA$ are exactly
$m$, and both the row and column margins of $\bfB$ are $mnp$.
Then, we can proceed
in two nearly-equivalent ways to perform MCA, corresponding operationally to a
standard CA on either the indicator matrix $\bfA$ or the Burt matrix
$\bfB$.  Forming the pseudo-residual matrices as before (Section \ref{sec:CA}) for each of
$\bfA$ and $\bfB$ and simplifying, we obtain
\begin{align}
  \bfZ_\bfA &
  =  \frac{1}{\sqrt{mn}}
  (\bfA-\bfone\bfp^T)\bfD_\bfp^{-1/2} \label{eq:acm}\\
  \bfZ_\bfB &
  = \frac{1}{mn}\bfD_\bfp^{-1/2}(\bfB-n\bfp\bfp^T)\bfD_\bfp^{-1/2} \nonumber
\end{align}
Since $\bfA^T\bfone\bfp^T = \bfp\bfone^T\bfone\bfp^T = n\bfp\bfp^T$, it follows immediately that $\bfZ_\bfB =
\bfZ_\bfA^T\bfZ_\bfA$, so that if $\bfZ_\bfA$ has SVD $\bfU \bfD
\bfV^T$, then $\bfZ_\bfB = \bfV \bfD^2 \bfV^T$, and we can recover the MCA coefficients
for the Burt-matrix form from the coefficients for the
indicator-matrix form (but not vice-versa).
We denote $\boldsymbol{\widehat\Gamma_{\text{MCA}}}$ to be
the MCA decomposition $\bfU_K\bfD_K\bfV_K^T$ of $\bfZ_\bfA$.

This specific choice of weighting and transformation in MCA implies that the principal components, denoted $\bf{f}_k$ for $k=1, ..., K$ satisfy:
$$ \bf{f}_k =\operatorname*{arg\,max}_{\bf{f}_k \in {\mathbb R}^{n}} \sum_{j=1}^{m}\eta^2(\bf{f}_k,X_{m})$$
with the constraint that $\bf{f}_k$  is orthogonal to $\bf{f}_k'$ for all $k'<k$ and $\eta^2$ the square of the correlation ratio (in an analysis of variance sense).  This formulation highlights that MCA can be seen as the counterpart of PCA for categorical data. In addition, the distances between the rows coincide with the  $\chi^2$ distances. 
MCA analysis mainly consists of interpreting the graphical outputs where rows are represented
with $\bfU \bfD \bfD_\bfp^{-1/2}$ and categories with $\bfV \bfD \bfD_\bfp^{-1/2}$ to identify rows with the same profile of response and association between categories. More properties are given in \cite{hussonjosse14}.

\subsection{MCA as One-Step Likelihood Estimates}\label{sec:connections}

Our main results in this section is to show that the low-rank
least-squares decompositions of the pseudo-residual matrices
$\bfZ_\bfA$ may be viewed as a one-step estimate for the
cognate model discussed in Section~\ref{sec:multiLogitBilinear} 
the \textit{multilogit-bilinear}. 

The rationale of the approach is the following one.
Each model represented in Table \ref{tab:methods}  is parametrized by $(\boldsymbol{\beta}, \boldsymbol{\Gamma})$, with the
constraint $\rank(\boldsymbol{\Gamma})\leq K$.  Maximizing $\ell(\boldsymbol{\beta},\boldsymbol{\Gamma};\bfX)$
is difficult owing to the non-convex constraint.
If instead we
Taylor expand $\ell$ around the {\em independence model}
$(\boldsymbol{\beta}_0,0)$ to obtain $\tilde\ell(\boldsymbol{\beta},\boldsymbol{\Gamma})$, a quadratic
function of its arguments, then maximizing the latter amounts to a
generalized singular value decomposition, which can be performed
efficiently.  Moreover, the generalized singular value problem is
precisely the one we solve to obtain the row and column
coordinates in  MCA.

\begin{table}
  \centering
  \begin{tabular}{c | c  c  c }
    Principal Component Method
    & Cognate Likelihood Model\\
    \hline
    CA
    & Log-bilinear Poisson \eqref{loglin}\\
    Indicator MCA
    & Multilogit-bilinear Model \eqref{eq:multiLogitBilinear}\\
  \end{tabular}
    \caption{Relationships between principal component methods for
      contingency tables and their cognate likelihood methods.  Each
      of the methods in the first row can be characterized as one-step
      estimates of the models in the second.}
      \label{tab:methods}
\end{table}

\paragraph{Quadratic Functions of a Matrix}

Let $\zeta = \binom{\beta}{\text{vec}(\Gamma)}$ denote the real vector of
all the model's parameters with $\zeta_0=\binom{\beta_0}{0}$, and
define the function $\tilde \ell$ to be the second-order Taylor
approximation:
\begin{equation*}
  \tilde\ell(\beta,\Gamma) = \ell(\zeta_0) +
  \ell'(\zeta_0)^T(\zeta - \zeta_0) +
  \frac{1}{2}(\zeta-\zeta_0)^T\ell''(\zeta_0)(\zeta-\zeta_0)
\end{equation*}
To begin, we establish a simple technical result that will arise in
the proof.  For matrices $\bfG,\bfH$ of the same shape, we use the notation
$\langle \bfG, \bfH\rangle = \Tr(\bfG^T\bfH) = \sum_{ij} G_{ij} H_{ij}$ to denote
the Frobenius inner product.

\begin{lemma}\label{lem:quadSol}
  Let $\bfG\in \R^{n\times n}, \bfH_1\in \R^{n\times n},
  \bfH_2\in \R^{m\times m}$, with $\bfH_1,\bfH_2 \succ 0$.  Then the problem
  \begin{equation}\label{eq:niceQuadForm}
    \argmax_{\boldsymbol{\Gamma}:\,\textnormal{rank}(\boldsymbol{\Gamma})\leq K} \langle \boldsymbol{\Gamma},
    \bfG\rangle - \frac{1}{2}\|\bfH_1 \boldsymbol{\Gamma} \bfH_2\|_F^2
  \end{equation}
  is solved by
  \begin{equation*}
   \boldsymbol{\Gamma}^* = \bfH_1^{-1}\left[\textnormal{SVD}_K(\bfH_1^{-1}\bfG\bfH_2^{-1})\right]\bfH_2^{-1}
  \end{equation*}
\end{lemma}

Lemma~\eqref{lem:quadSol} proven in Appendix \ref{lem:quadSolapp} is easy but vital, since it
gives us a target to aim for when we
construct Newton approximations to the log-likelihoods of interest.
The first-order term in our Taylor expansion is necessarily of the
form $\langle\boldsymbol{\Gamma}, \bfG\rangle$, where $G_{ij}$ is simply the gradient with
respect to entry  $\Gamma_{ij}$.  Hence, if we could only show the second
derivative term is of the form $-\frac{1}{2}\|\bfH_1 \boldsymbol{\Gamma} \bfH_2\|_F^2$,
then our problem would reduce to a generalized singular value decomposition.


\paragraph{MCA and the Multilogit-Bilinear Model}

\begin{theorem} \label{theo:maintheo}
  The one-step likelihood estimate for the model
  \eqref{eq:multiLogitBilinear} with rank constraint $K$,
  obtained by expanding around the independence model $(\boldsymbol{\beta}_0=\log
  \boldsymbol{p},\boldsymbol{\Gamma}_0=0)$, is
  $(\boldsymbol{\beta}_0,\boldsymbol{\widehat\Gamma_{\text{MCA}}})$.
\end{theorem}

 The log-likelihood for the model \eqref{eq:multiLogitBilinear} is
  \begin{equation*}
   \ell(\boldsymbol{\beta}, \boldsymbol{\Gamma};\bfA) = \beta^j(a_{ij}) + \Gamma_i^j(a_{ij})
    - \log\left(\sum_{c=1}^{C_j} e^{\beta^j(c) + \Gamma_i^j(c)} \right)
  \end{equation*}
It is easy to show (see Appendix \ref{sec:prooftheo}), that that the total contribution in the second-order Taylor approximation evaluating  at $(\boldsymbol{\beta}_0,0)$ of the linear term is $\langle
  \boldsymbol{\Gamma}, \bfA - \bfone\bfp^T\rangle$ and that the total contribution of the second derivatives in $\boldsymbol{\Gamma}$ is  $- \frac{1}{2}\|\boldsymbol{\Gamma} \bfD_p^{1/2}\|_F^2$. Thus, using Lemma \ref{lem:quadSol}, the solution is given by the rank $K$ SVD of $(\bfA-\bfone\bfp^T)\bfD_\bfp^{-1/2}$  which is precisely the SVD performed in MCA (equation \eqref{eq:acm}).

\section{Empirical Comparison to MCA}  \label{sec:simulations}

In Section \ref{sec:connections}, we showed that the parameters estimated by MCA can be seen as providing an approximation of the parameters of the \textit{multilogit-bilinear model} when the interaction is low. We assess empirically Theorem 2 in a simulation study where the data are simulated according to the \textit{multilogit-bilinear model} varying several parameters: 
\begin{itemize}
\item the number of individuals $n$ (50, 100, 300), the number of variables $m$ (20, 100, 300). The number of categories per variable is set to 3.  
\item the number of terms of the interaction $K$ (2, 6)
\item the ratio of the first singular value to the second
singular value $(d_1/d_2)$ (2, 1). When $K$ is
greater than 2, the subsequent singular values
are of the same order of magnitude.
\item the strength of the interaction (low, strong)
\end{itemize}
More precisely:
\begin{align*}
    \tilde \bfu_i &\sim \mathcal{N}_K\left(0,\begin{pmatrix} d_1 & 0 \\ 0 & d_K\end{pmatrix}\right)\\
    \tilde \bfv_j(c) &\sim \mathcal{N}_K\left(0,\begin{pmatrix} d_1 & 0 \\ 0 & d_K\end{pmatrix}\right)\\
    \theta_{ij}^c &= -\frac{1}{2}\|\tilde \bfu_i-\tilde \bfv_j(c)\|^2\\
    \P(x_{ij}=c) &\propto e^{\theta_{ij}^c}
  \end{align*}
The strength of the interaction is controlled by multiplying the singular values by a term equal to 0.1 (low) or 1 (strong).
To estimate the parameters of the \textit{multilogit model} \eqref{eq:multiLogitBilinear} we use the majorization algorithm suggested in \cite{groen} implemented in the R package \textit{mmca} \citep{pat} without any penalty. We perform MCA using the R package \textit{FactoMineR} \citep{facto}.
A representative extract of the results is given in Table \ref{tab:ressimu}. The standard deviations of the MSEs are very small  and vary from the order of $10^{-5}$  to $10^{-3}$ (for small sample size). Thus, the MSEs can be directly analyzed to compare the estimators.

\begin{table}[ht]
\centering
\begin{tabular}{rrrrrrrr} 
  \hline
 & $n$ & $p$  & rank & ratio & strength & model & MCA \\ 
  \hline
1& 50& 20 & 2 & 1 & 0.1 & 0.044 & \textbf{0.035} \\ 
  2 & 50& 20 & 2  & 1 & 1 & 0.020 & 0.045 \\ 
  3 &50& 20 & 2 & 2 & 0.1& 0.048 & \textbf{0.036} \\ 
  4 & 50& 20 & 2 & 2 & 1 & \textbf{0.0206} & 0.042\\ 
  5 & 50& 20 & 6 & 1 & 0.1 & 0.111 & \textbf{0.064} \\ 
   6& 50& 20 & 6 & 1 & 1 & 0.045 & \textbf{0.026} \\ 
   7& 50& 20 & 6 & 2 & 0.1 & 0.115 (0.028)& \textbf{0.071} \\ 
   8 &50& 20 & 6  & 2 & 1& \textbf{0.032} & 0.051 \\

 9& 300 & 100& 2 & 1 & 0.1 & \textbf{0.005} & \textit{0.006}\\ 
  10 & 300 & 100& 2 & 1 & 1 & \textbf{0.004} &  0.042\\ 
  11 & 300 & 100& 2& 2 & 0.1& \textbf{0.0047} &  \textit{0.005}\\ 
  12 & 300 & 100&  2 & 2 & 1 & \textbf{0.0037} (0.00369)&  0.040 \\
 13& 300 & 300 & 2 & 1 & 0.1 & \textbf{0.003} & \textit{0.004} \\ 
   14& 300 & 300  & 2 & 1 & 1 & \textbf{0.002}  & 0.039 \\ 
   15& 300 & 300 & 2 & 2 & 0.1 & \textbf{0.003} & \textit{0.004} \\ 
   16& 300 & 300  & 2 & 2 & 1 & \textbf{0.002} & 0.039 \\ 

17& 300 & 100& 6 & 1 & 0.1& 0.019 & \textit{\textbf{0.015}}  \\ 
  18 & 300 & 100  & 6 & 1 & 1& \textbf{0.011} & 0.023 \\ 
   19& 300 & 100  & 6 & 2 & 0.1& 0.018 (0.010) & \textit{\textbf{0.017}} \\ 
   20& 300 & 100 & 6 &2& 1& \textbf{0.010} & 0.056 \\ 
   21& 300 & 300& 6 & 1& 0.1& 0.011 & \textit{\textbf{0.008}}  \\ 
   22& 300 & 300 & 6 & 1& 1 & \textbf{0.006} & 0.022  \\ 
   23& 300 & 300 & 6 & 2& 0.1 & \textbf{0.009} & \textit{0.012}  \\ 
   24& 300 & 300 & 6 & 2& 1 & \textbf{0.006} & 0.061 \\ 

\end{tabular} 
\caption{Root mean square error (RMSE) between true probabilities and estimated ones by the \textit{multilogit model} and MCA. In bold the smallest MSEs; in italic when MCA results are of the same order of magnitude than the ones obtained by the model. For cases, 7, 12 and 19, the error obtained when using a penalized version of the likelihood to estimate the model parameters \citep{groen} is also given in brackets.} \label{tab:ressimu}
\end{table}

\begin{figure}[!hbt]
\begin{center}
\includegraphics[scale=0.8]{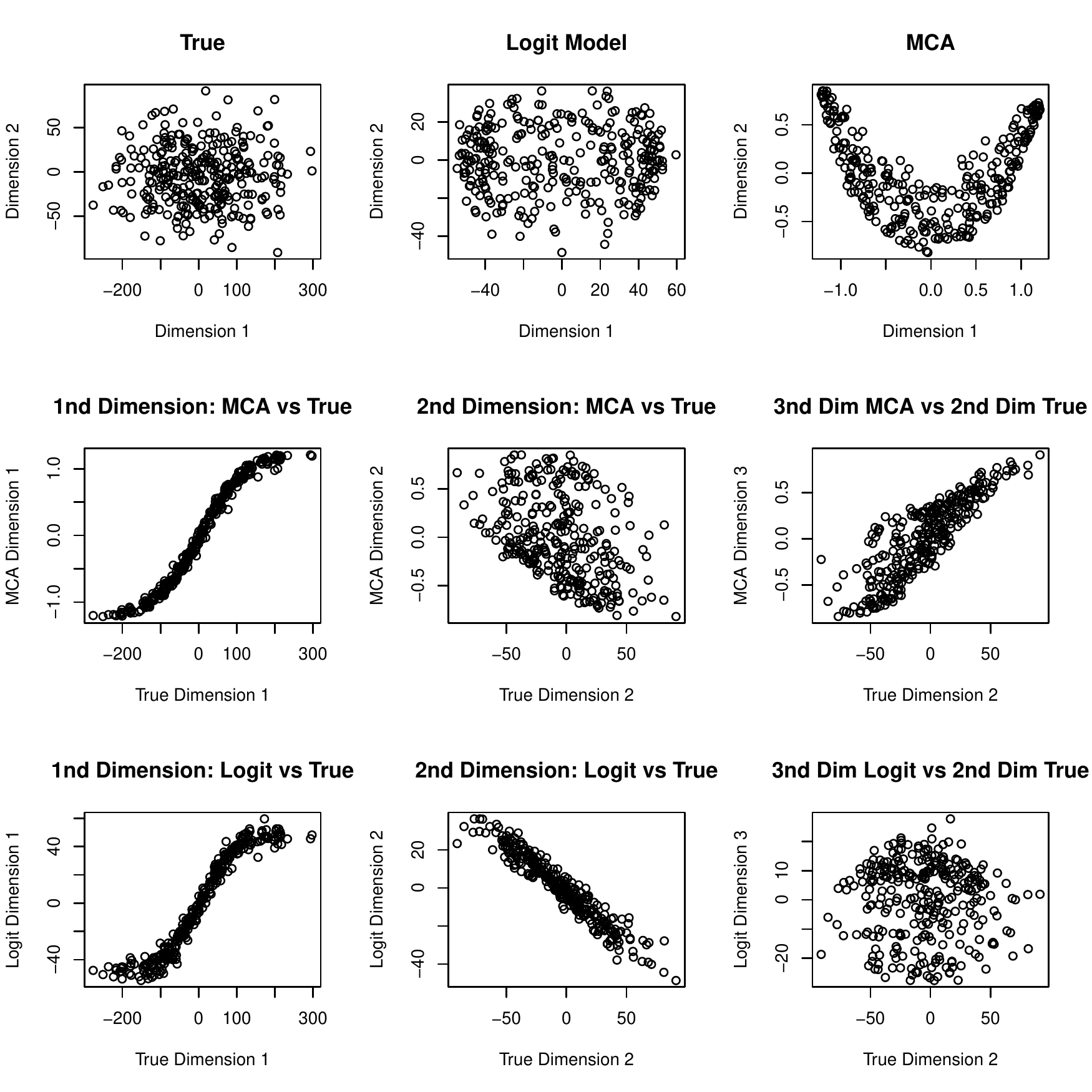}
\caption{Estimation of the parameters by MCA and by the majorization algorithm in a case with a strong interaction and first dimension. The parameters are $n=300, p=100, r=2$.}
\label{fig:strong}
\end{center}
\end{figure}

As expected, when the strength of the interaction is low (0.1), both methods agree: the parameters estimated by MCA and by the majorization algorithm are very correlated to the true parameters whatever the scenario and the MSEs are of the same order of magnitude. Thus, MCA is a straightforward alternative, computationally fast and easy to run, to accurately estimate the parameters of the \textit{multilogit-bilinear model}. 
When the signal is strong different patterns occur.  Figure \ref{fig:strong} illustrates a case with a strong first dimension of variability. 
The majorization algorithm recovers well the true dimensions whereas MCA exhibits an horseshoe effect. Its second dimension
can be viewed as an artifact. Consequently, MCA does not seem appropriate to estimate the parameters of  \textit{multilogit-bilinear} model. Nevertheless, one can note that the signal is not completely lost since MCA third dimension of variability corresponds to the true second one. 
On all the experiments carried out, we also saw situations where both MCA and the majorization algorithm exhibit an horshoe effect \citep{Gutt53}. It could be interesting to investigate more the understanding of this behavior in the context of the multilogit model as it was done in \citet{Bach94}, \citet{deLeeuw07} and \citet{Diaco08} in the framework of CA and multidimensional scaling.
Nevertheless, when the signal is strong even if MCA is less appropriate to estimate the parameters (the MSEs are around 10 times larger),  we feel after inspecting many plots, that the approximation is accurate enough and will often lead to the same interpretation of the results. This is in agreement with what was observed in CA by \citet{Heijden94}.  
Finally, it may seem surprising to see that MCA can provide better estimates than the model ones. This occurs in what can be considered as difficult settings with small $n$ and $p$ and/or noisy data where the strength of the relationship is weak and/or the rank $K$ is large (cases 1, 3, 5, 6, 7). In such situations, the majorization algorithm may encounter difficulties and it is necessary to resort to regularization. If we use a regularized version with the amount of shrinkage determined by cross-validation \citep{groen}, the error for case 7 is then equal to 0.028 instead of 0.115 and improves on MCA. On the contrary, the impact of regularization is less crucial in "easy" frameworks (case 12).
The results are reproducible with the R code provided on a github repository \citep{git}.

 
\section{Discussion}

Theoretical connections between CA and the \textit{log-bilinear} model were suggested in the literature but it was lacking for MCA. In this paper,  we showed that MCA can be seen as a  linearized estimate of the parameters of  the \textit{multinomial logit bilinear  model}. Thus, MCA can be used as a proxy to estimate the model's parameters. In a simulation study, we showed that  MCA is particularly well fitted  in regimes with small interaction and often provides a good approximation in the other cases. 
These tight connections allow a better understanding of both models and exploratory methods and going back and forth is part of the process to enhance the comprehension of the approaches and give them a larger scope. 

For instance, regularization in the \textit{multi-logit} model is crucial for better estimation in noisy schemes but the practice is less common in MCA (see \cite{Takane06} and \cite{Josse12} in the framework of missing values). The established relationship between MCA and the \textit{multi-logit model} greatly encourages to regularize MCA to tackle overfitting issues. In a similar way, graphical outputs are at the core of MCA analysis and almost never used within the probabilistic framework. The experience in the graphical representations in MCA should be used to display the results of the \textit{multi-logit model} to enhance the interpretation of the results. Finally, we should also mention that MCA is a very powerful way to predict missing values \citep{Audi15}, the connection with the model gives more rational and strengthens this good behavior.

We finish by discussing some opportunities for further research. 
A natural area that should be investigated is to extend mixtures of PCA \citep{mppca} to categorical data with mixtures of MCA. Since no model was associated to MCA, this mixture model was never considered. Such an approach would allow to get rid of the strong hypothesis of independence between categorical variables within a cluster that is often assumed. 
Another point that can be considered is the analysis of mixed data (both continuous and categorical data) with the method factorial analysis for mixed data (FAMD) \citep{Escofier79, Kiers91} and  data structured in groups of variables with methods such as multiple factor analysis (MFA) \citep{Pages14}.  The extension of the theoretical connections between a  cognate model and these exploratory methods is not straightforward since specific weightings are applied to balance the influence of variables of different nature. Finally, no method is available to select the rank in MCA and few methods are available to get confidence areas around the results. Using model selections criteria for the \textit{multinomial logit bilinear model} should at least give hints on the number of relevant dimensions which is crucial for the MCA analysis.  This point is definitively a worthwhile enterprise.


\bibliographystyle{spbasic}  
\bibliography{jossee}   

\section{Appendix: Proofs}

\subsection{Proof of Lemma \ref{lem:quadSol} \label{lem:quadSolapp}}

\begin{proof}
  Change variables to $\widetilde \Gamma = H_1\Gamma H_2$ and complete
  the square with constant $g=\frac{1}{2}\|H_1^{-1}G
  H_2^{-1}\|_F^2$.  Then we obtain
  \begin{align*}
    \langle \Gamma,
    G\rangle - \frac{1}{2}\|H_1 \Gamma H_2\|_F^2- g
    &= \langle H_1^{-1}\widetilde\Gamma H_2^{-1}, G\rangle \;-\;
    \frac{1}{2}\|\widetilde \Gamma\|_F^2
    \;-\; \frac{1}{2}\|H_1^{-1}G H_2^{-1}\|_F^2\\
    &=
    \langle\widetilde\Gamma,  H_1^{-1}GH_2^{-1}\rangle \;-\;
    \frac{1}{2}\|\widetilde \Gamma\|_F^2
    \;-\; \frac{1}{2}\|H_1^{-1}G
    H_2^{-1}\|_F^2\\
    &= -\frac{1}{2}\|\widetilde\Gamma - H_1^{-1}GH_2^{-1}\|_F^2
  \end{align*}
  Solving for $\widetilde \Gamma$ amounts to a rank-$K$ SVD,
  and we transform back to obtain the result.
\end{proof}

\subsection{Proof of Theorem \ref{theo:maintheo}} \label{sec:prooftheo}

\begin{proof}
  The log-likelihood for the model \eqref{eq:logisticBilinear} is
  \begin{equation*}
    \ell(\boldsymbol{\beta}, \boldsymbol{\Gamma};\bfA) = \beta^j(a_{ij}) + \Gamma_i^j(a_{ij})
    - \log\left(\sum_{c=1}^{C_j} e^{\beta^j(c) + \Gamma_i^j(c)} \right)
  \end{equation*}
  Differentiating once with respect to $\Gamma_i^j(c)$, we obtain
  \begin{equation}\label{eq:MCAdG}
    \pardd{\ell}{\Gamma_i^j(c)} = 1_{x_{ij}=c} -
    \frac{e^{\beta^j(c) + \Gamma_i^j(c)}}
    {\sum_{c'=1}^{C_j} e^{\beta^j(c') + \Gamma_i^j(c')}}
  \end{equation}
  Evaluating \eqref{eq:MCAdG} at $(\beta_0,0)$ gives
  \begin{equation}\label{eq:MCAdGres}
    A_i^j(c) - p^j(c),
  \end{equation}
  so that the total contribution of the linear term is $\langle
  \Gamma, A - 1p^T\rangle$.

  Differentiating \eqref{eq:MCAdG} with respect to $\beta^{j'}(c')$, and
  evaluating at $(\beta_0,0)$ gives
  \begin{equation*}
    \pardd{\ell}{\Gamma_i^j(c)\partial \beta^{j'}(c')}
    = \left\{
      \begin{matrix}
        p^j(c)p^j(c') - p^j(c)1_{c=c'} & j = j'\\
        0 & \text{ o.w. }\\
      \end{matrix}
    \right.
  \end{equation*}
  Thus, the total contribution to $\tilde\ell$
  of crossed partials involving
  $\beta^{j}(c')$ is
  \begin{align*}
    \frac{1}{2}(\beta^j(c)-\beta_0^j(c)) \sum_{i,c'}
    \pardd{\ell}{\Gamma_i^j(c')\partial \beta^j(c')} \Gamma_i^j(c')
    &= \frac{1}{2}(\beta^j(c)-\beta_0^j(c)) p^j(c) \sum_{i,c'} \Gamma_i^j(c')
    (p^j(c')-1_{c=c'})\\
    &= \frac{1}{2}(\beta^j(c)-\beta_0^j(c)) p^j(c) \sum_{i} -\Gamma_i^j(c)
  \end{align*}

  Differentiating \eqref{eq:MCAdG} with respect to
  $\Gamma_{i'}^{j'}(c')$ and evaluating at $(\beta_0,0)$ gives
  \begin{equation*}
    \pardd{\ell}{\Gamma_i^j(c)\partial \Gamma_{i'}^{j'}(c')} = \left\{
      \begin{matrix}
        p^j(c)p^j(c') - p^j(c)1_{c=c'} & j = j', i=i'\\
        0 & \text{ o.w. }\\
      \end{matrix}
    \right.
  \end{equation*}

  The total contribution of the second derivatives in $\Gamma$ is then
  \begin{align*}
    -\frac{1}{2}\sum_{i,j,c} \Gamma_i^j(c)^2p^j(c) + \frac{1}{2}\sum_{i,j,c,c'}
    p^j(c)\Gamma_i^j(c)p^j(c')\Gamma_i^j(c')
    &= -\frac{1}{2}\|\Gamma D_p^{1/2}\|_F^2 +
    \frac{1}{2}\sum_{i,j} \left(\sum_c \Gamma_i^j(c)p^j(c)\right)^2\\
    &= - \frac{1}{2}\|\Gamma D_p^{1/2}\|_F^2
  \end{align*}

  Overall, then, we have
  \begin{align*}
    \tilde\ell(\beta,\Gamma) = \langle \Gamma, A - 1p^T\rangle
    - \frac{1}{2} \|\Gamma D_p^{1/2}\|_F^2
    - \frac{1}{2} 1^T\Gamma D_p (\beta-\beta_0)
  \end{align*}

\end{proof}

\vspace{0.5cm}

\subsection*{Acknowledgment}

\noindent We thank Antoine de Falguerolles for interesting discussions and his "glim" view on exploratory methods.

\end{document}